\documentclass[preprint,journal]{vgtc}       

\ieeedoi{10.1109/TVCG.2020.3030378}

\usepackage{microtype}                 
\PassOptionsToPackage{warn}{textcomp}  
\usepackage{textcomp}                  
\usepackage{mathptmx}                  
\usepackage{times}                     
\usepackage{cite}                      
\usepackage{tabu}                      
\usepackage{booktabs}                  

\usepackage[normalem]{ulem}
\usepackage{enumitem}
\usepackage{calc}
\usepackage{textcomp}
\usepackage{graphicx}
\usepackage{soul}
\usepackage[dvipsnames]{xcolor}
\usepackage{tcolorbox}
\usepackage[frozencache,cachedir=.]{minted}
\usepackage{listings}
\usepackage[font={color={black}}]{caption}
\usepackage{fontawesome}
\usepackage{multirow}
\usepackage{tcolorbox}
\usepackage{hyperref}

\newcommand{\add}[1]{\textcolor{black}{#1}}

\newcommand{\canremove}[1]{\textcolor{black}{#1}}

\definecolor{customLightGreen}{HTML}{D9EAD3}
\definecolor{customCodeBlue}{HTML}{214A87}
\definecolor{customGreen}{HTML}{54AA54}
\definecolor{customCodeOrange}{HTML}{C55A11}
\definecolor{darkteal}{HTML}{004B53}
\definecolor{darkred}{HTML}{A50021}
\definecolor{urlblue}{HTML}{067DE9}

\definecolor{explicitBlack}{HTML}{404040}
\definecolor{ambiguousGray}{HTML}{DEDEDE}

\definecolor{explicitBlue}{HTML}{0070C0}
\definecolor{ambiguousBlue}{HTML}{CCECFF}
\definecolor{implicitBlue}{HTML}{002060}

\newtcbox{\expliciptCapTag}{nobeforeafter, colback=explicitBlue, boxrule=0.5pt, arc=1pt, colframe=explicitBlue, boxsep=0pt,left=2pt,right=2pt,top=1.75pt,bottom=1.5pt,tcbox raise base}
\newtcbox{\ambiguousCapTag}{nobeforeafter, colback=ambiguousBlue, boxrule=0.5pt, arc=1pt, boxsep=0pt,left=2pt,right=2pt,top=1.75pt,bottom=1.5pt,tcbox raise base}
\newtcbox{\implicitCapTag}{nobeforeafter, colback=white, colframe=implicitBlue, boxrule=0.5pt, arc=1pt, boxsep=0pt,left=2pt,right=2pt,top=1.75pt,bottom=1.5pt,tcbox raise base}

\definecolor{key0Color}{HTML}{5FaD56}
\definecolor{key1Color}{HTML}{ca7611}
\definecolor{key2Color}{HTML}{ca7611}
\definecolor{valueColor}{HTML}{1768AC}

\newcommand{\variable}[1]{\textcolor{customCodeBlue}{\texttt{\textbf{#1}}}}
\newcommand{\function}[1]{\colorbox{Gray!10}{\texttt{\textbf{\small{#1}}}}}
\newcommand{\param}[1]{\texttt{\textbf{#1}}}
\newcommand{\attrType}[1]{\textit{#1}}

\newcommand{\implementationchallenge}[1]{\colorbox{darkteal}{\textcolor{white}{\small{#1}}}}

\newcommand{\key}[1]{\textcolor{key0Color}{#1}}
\newcommand{\sKey}[1]{\textcolor{key1Color}{\texttt{#1}}}
\newcommand{\ssKey}[1]{\textcolor{key2Color}{\texttt{#1}}}
\newcommand{\val}[1]{\textcolor{valueColor}{#1}}

\newcommand{\systemURL}{\href{https://nl4dv.github.io/nl4dv/}{\textcolor{urlblue}{\texttt{https://nl4dv.github.io/nl4dv/}}}}

\onlineid{0}

\vgtcinsertpkg

\title{
NL4DV: A Toolkit for Generating Analytic Specifications\\ for Data Visualization from Natural Language Queries
}

\author{Arpit Narechania*, Arjun Srinivasan*, and John Stasko}
\authorfooter{
\item
 Arpit Narechania, Arjun Srinivasan, and John Stasko are from the Georgia Institute of Technology, Atlanta, GA (USA). E-mail: \{arpitnarechania, arjun010\}@gatech.edu, stasko@cc.gatech.edu.
\item
 *Authors contributed equally.
}

\shortauthortitle{Biv \MakeLowercase{\textit{et al.}}: Global Illumination for Fun and Profit}

\abstract{
Natural language interfaces (NLIs) have shown great promise for visual data analysis, allowing people to flexibly specify and interact with visualizations.
However, developing visualization NLIs remains a challenging task, requiring low-level implementation of natural language processing (NLP) techniques as well as knowledge of visual analytic tasks and visualization design.
We present NL4DV, a toolkit for natural language-driven data visualization.
NL4DV is a Python package that takes as input a tabular dataset and a natural language query about that dataset.
In response, the toolkit returns an analytic specification modeled as a JSON object containing data attributes, analytic tasks, and a list of Vega-Lite specifications relevant to the input query.
In doing so, NL4DV aids visualization developers who may not have a background in NLP, enabling them to create new visualization NLIs or incorporate natural language input within their existing systems.
We demonstrate NL4DV's usage and capabilities through four examples: 1) rendering visualizations using natural language in a Jupyter notebook, 2) developing a NLI to specify and edit Vega-Lite charts, 3) recreating data ambiguity widgets from the DataTone system, and 4) incorporating speech input to create a multimodal visualization system.
}


\CCScatlist{ 
 \CCScat{K.6.1}{Management of Computing and Information Systems}%
{Project and People Management}{Life Cycle};
 \CCScat{K.7.m}{The Computing Profession}{Miscellaneous}{Ethics}
}

\teaser{
  \centering
  \includegraphics[width=.85\linewidth]{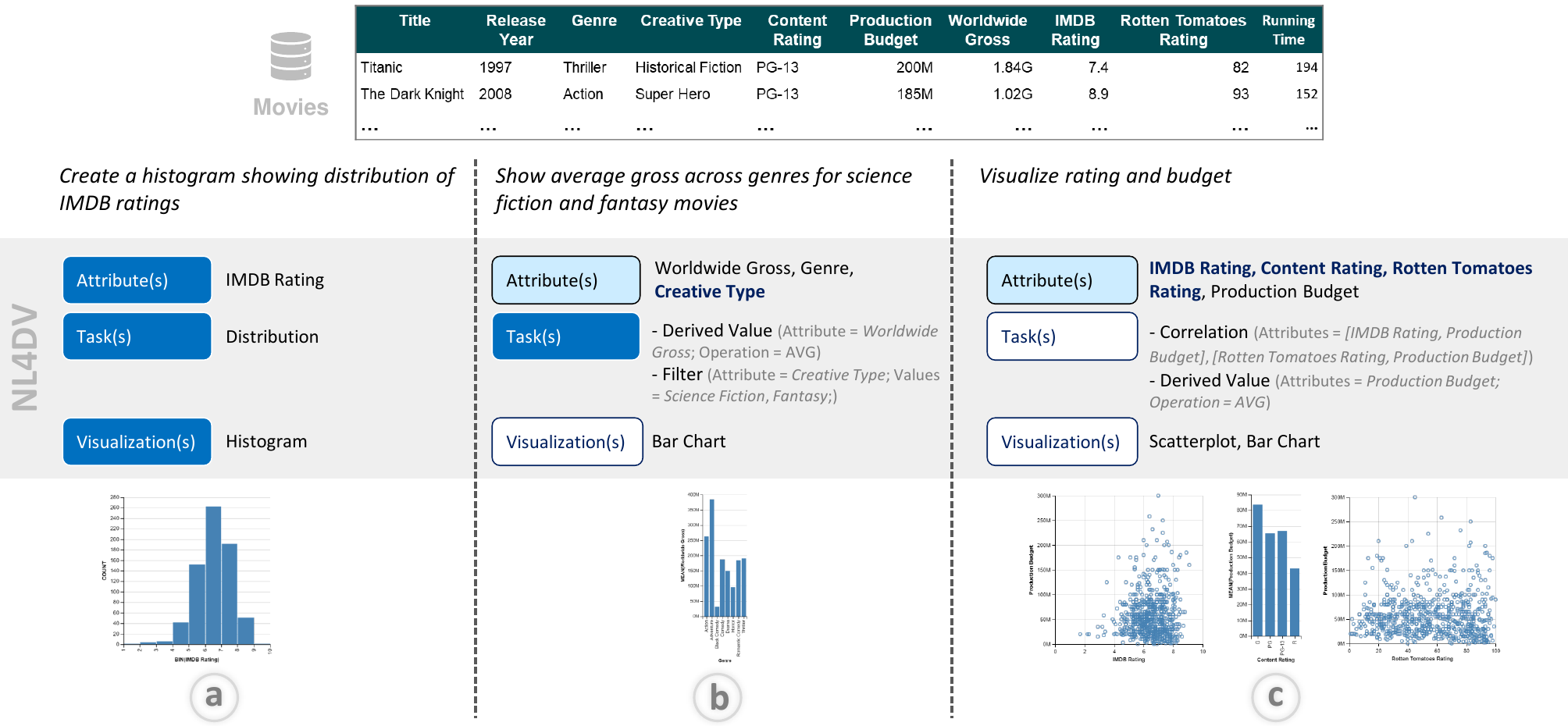}
  \vspace{-.5em}
  \caption{
  Examples illustrating the flexibility of natural language queries for specifying data visualizations.
  NL4DV processes all three query variations, inferring
  \expliciptCapTag{\textcolor{white}{explicit}},
  \ambiguousCapTag{\textcolor{black}{partially explicit}} or \ambiguousCapTag{\textcolor{black}{ambiguous}},
  and \implicitCapTag{\textcolor{implicitBlue}{implicit}}
  references to attributes, tasks, and visualizations.
  The corresponding visualizations suggested by NL4DV in response to the individual queries are also shown.
  }
  \label{fig:teaser}
  \vspace{-.5em}
}

\begin{document}



\firstsection{Introduction}
\maketitle

Natural language interfaces (NLIs) for visualization are becoming increasingly popular in both academic research (e.g.,~\cite{gao2015datatone,setlur2016eviza,hoque2018applying,srinivasan2018orko,yu2019flowsense}) as well as commercial software~\cite{mspowerbi,tableauaskdata}.
At a high-level, visualization NLIs allow people to pose data-related queries and generate visualizations in response to those queries.
To generate visualizations from natural language (NL) queries, NLIs first model the input query in terms of \textit{data attributes} and \textit{low-level analytic tasks}~\cite{amar2005low,chen2009toward} (e.g., filter, correlation, trend).
Using this information, the systems then determine which visualizations are most suited as a response to the input query. 
While NLIs provide flexibility in posing data-related questions, inherent characteristics of NL such as ambiguity and underspecification make implementing NLIs for data visualization a challenging task.


To create visualization NLIs, besides implementing a graphical user interface (GUI) and rendering views, visualization system developers must also implement a natural language processing (NLP) module to interpret queries.
Although there exist tools to support GUI and visualization design (e.g., D3.js~\cite{bostock2011d3}, Vega-Lite~\cite{satyanarayan2016vega}), developers currently have to implement custom modules for query interpretation.
However, for developers without experience with NLP techniques and toolkits (e.g., NLTK~\cite{loper2002nltk}, spaCy~\cite{honnibal2017spacy}), implementing this pipeline is non-trivial, requiring them to spend significant time and effort in learning and implementing different NLP techniques.


Consider the spectrum of queries in Figure~\ref{fig:teaser} issued to create visualizations in the context of an IMDb movies dataset
with different attributes including the {\small{\faHashtag}}~Worldwide Gross, {\small{\faFont}}~Genre, and {\small{\faCalendar}}~Release Year, among others (for consistency, we use this movies dataset for examples throughout this paper).
The query \textit{``Create a histogram showing distribution of IMDB ratings"} (Figure~\ref{fig:teaser}a) explicitly refers to a data attribute (\textit{IMDB Rating}), a low-level analytic task (\textit{Distribution}), and requests a specific visualization type (\textit{Histogram}).
This is an ideal interpretation scenario from a system standpoint since the query explicitly lists all components required to generate a visualization.

On the other hand, the second query \textit{``Show average gross across genres for science fiction and fantasy movies"} (Figure~\ref{fig:teaser}b) does not explicitly state the visualization type or the attribute \textit{Creative Type}.
Instead, it explicitly references the attributes \textit{Worldwide Gross} and \textit{Genre} through `gross' and `genres', and implicitly refers to the \textit{Creative Type} through the values `science fiction' and `fantasy'.
Furthermore, by specifying data values for the \textit{Creative Type} attribute and the word `average,' the query also mentions two intended analytic tasks: \textit{Filtering} and computing \textit{Derived Values}, respectively.
This second query is more challenging since it requires the system to implicitly infer one of the attributes and then determine the visualization type based on the identified attributes and tasks.

Finally, the third query \textit{``Visualize rating and budget"} (Figure~\ref{fig:teaser}c) is even more challenging to interpret since it neither explicitly states the desired visualization type nor the intended analytic task.
Furthermore, while it explicitly references one attribute (\textit{Production Budget} through `budget'), the reference to the second attribute is ambiguous (`rating' can map to \textit{IMDB Rating}, \textit{Content Rating}, or \textit{Rotten Tomatoes Rating}).

To accommodate such query variations, visualization NLIs employ sophisticated NLP techniques (e.g., dependency parsing, semantic word matching) to identify relevant information from the query and build upon visualization concepts (e.g., analytic tasks) and design principles (e.g., choosing graphical encodings based on attribute types) to generate appropriate visualizations.
For instance, given the query in Figure~\ref{fig:teaser}b, after detecting the data attributes and analytic tasks, a visualization NLI should select a visualization (e.g., bar chart) that is well-suited to support the task of displaying \textit{Derived Values} (average) for \textit{Worldwide Gross} (a quantitative attribute) across different \textit{Genre}s (a nominal attribute).
Similarly, in the scenario in Figure~\ref{fig:teaser}c, a NLI must first detect ambiguities in the input query attributes, determine the visualizations suited to present those attribute combinations (e.g., \textit{scatterplot} for two quantitative attributes), and ultimately infer the analytic tasks based on those attributes and visualizations (e.g., a scatterplot may imply the user is interested in finding \textit{correlations}).

To support prototyping NLIs for data visualization, we contribute the \textit{Natural Language-Driven Data Visualization} (\textit{NL4DV}) toolkit.
NL4DV is a Python package that developers can initialize with a tabular dataset.
Once initialized, NL4DV processes subsequent NL queries about the dataset, inferring data attributes and analytic tasks from those queries.
Additionally, using built-in mappings between attributes, tasks, and visualizations, NL4DV also returns an ordered list of Vega-Lite specifications relevant to those queries.
By providing a high-level API to translate NL queries to visualizations, NL4DV abstracts out the core task of interpreting NL queries and provides task-based visualization recommendations as plug-and-play functionality.
Using NL4DV, developers can create new visualization NLIs as well as incorporate NL querying capabilities into their existing visualization systems.

In this paper, we discuss NL4DV's design goals and describe how the toolkit infers data attributes, analytic tasks, and visualization specifications from NL queries.
Furthermore, we formalize the inferred information into a JSON-based analytic specification that can be programmatically parsed by visualization developers.
Finally, through example applications, we showcase how this formalization can help: 1) implement visualization NLIs from scratch, 2) incorporate NL input into an existing visualization system, and 3) support visualization specification in data science programming environments.

\canremove{To support development of future systems, we also provide NL4DV and the described applications as open-source software available at: \systemURL}


\section{Related Work}


\subsection{Natural Language Interfaces for Data Visualization}

In 2001, Cox et al.~\cite{cox2001multi} presented an initial prototype of a NLI that supported using well-structured commands to specify visualizations.
Since then, given the advent of NL understanding technology and NLIs for databases (e.g.,~\cite{popescu2003towards,blunschi2012soda,li2014nalir,pasupat2015compositional,zhongSeq2SQL2017,wang2018robust,he2019x,herzig2020tapas}), there has been a surge of NLIs for data visualization~\cite{sun2010articulate,gao2015datatone,setlur2016eviza,hoque2018applying,srinivasan2018orko,yu2019flowsense,srinivasan2020inchorus,setlur2019inferencing,srinivasan2020interweaving,kumar2016towards,kassel2018valletto,kim2020answering}, especially in recent years.
Srinivasan and Stasko~\cite{srinivasan2017natural} summarize a subset of these NLIs, characterizing systems based on their supported capabilities including \textit{visualization-focused} capabilities (e.g., specifying or interacting with visualizations), \textit{data-focused} capabilities (e.g., computationally answering questions about a dataset), and \textit{system control-focused} capabilities (e.g., augmenting graphical user interface actions like moving windows with NL).
Along these lines, NL4DV's current focus is primarily to support \textit{visualization specification}.
With this scope in mind, below we highlight systems that serve as the motivation for NL4DV's development and are most relevant to our work.

Articulate~\cite{sun2010articulate} is a visualization NLI that allows people to generate visualizations by deriving mappings between tasks and data attributes in user queries.
DataTone~\cite{gao2015datatone} uses a combination of lexical, constituency, and dependency parsing to let people specify visualizations through NL.
Furthermore, detecting ambiguities in the input query, DataTone leverages mixed-initiative interaction to resolve these ambiguities through GUI widgets such as dropdown menus.
FlowSense~\cite{yu2019flowsense} uses semantic parsing techniques to support NL interaction within a dataflow system, allowing people to specify and connect components without learning the intricacies of operating a dataflow system.
Eviza~\cite{setlur2016eviza} incorporates a probabilistic grammar-based approach and a finite state machine to allow people to interact with a given visualization.
Extending Eviza's capabilities and incorporating additional pragmatics concepts, Evizeon~\cite{hoque2018applying} allows both specifying and interacting with visualizations through standalone and follow-up utterances.
The ideas in Eviza and Evizeon were also used to design the Ask Data feature in Tableau~\cite{tableauaskdata}.
Ask Data internally uses Arklang~\cite{setlur2019inferencing}, an intermediate language developed to describe NL queries in a structured format that Tableau's VizQL~\cite{stolte2002polaris} can parse to generate visualizations.

The aforementioned systems all present different interfaces and capabilities, supporting NL interaction through grammar- and/or lexical-parsing techniques.
A commonality in their underlying NLP pipeline, however, is the use of data attributes and analytic tasks (e.g., correlation, distribution) to determine user intent for generating the system response.
Building upon this central observation and prior system implementations (e.g., string similarity metrics and thresholds~\cite{gao2015datatone,setlur2016eviza,srinivasan2018orko}, parsing rules~\cite{gao2015datatone,yu2019flowsense,kim2020answering}), NL4DV uses a combination of lexical and dependency parsing-based techniques to infer attributes and tasks from NL queries.
However, unlike previous systems that implement custom NLP engines and languages that translate NL queries into system actions, we develop NL4DV as an interface-agnostic toolkit.
In doing so, we formalize attributes and tasks inferred from a NL query into a structured JSON object that can be programmatically parsed by developers.


\subsection{Visualization Toolkits and Grammars}

Fundamentally, our research falls under the broad category of user interface toolkits~\cite{myers2000past,olsen2007evaluating,ledo2018evaluation}.
As such, instead of presenting a single novel technique or interface, we place emphasis on reducing development viscosity, lowering development skill barriers, and enabling replication and creative exploration.
Within visualization research, there exist a number of visualization toolkits with similar goals that particularly focus on easing development effort for specifying and rendering visualizations.
Examples of such toolkits include Prefuse~\cite{heer2005prefuse}, Protovis~\cite{bostock2009protovis}, and D3~\cite{bostock2011d3}.
With the advent of visualizations on alternative platforms like mobile devices and AR/VR, a new range of toolkits are also being created to assist visualization development on these contemporary platforms.
For instance, EasyPZ.js~\cite{schwab2019easypz} supports incorporating navigation techniques (pan and zoom) in web-based visualizations across both desktops and mobile devices.
Toolkits like DXR~\cite{sicat2018dxr} enable development of expressive and interactive visualizations in Unity~\cite{unity} that can be deployed in AR/VR environments.
NL4DV extends this line of work on toolkits for new modalities and platforms by making it easier for visualization system developers to interpret NL queries without having to learn or implement NLP techniques.

Besides toolkits that aid programmatically creating visualizations, researchers have also formulated visualization grammars that provide a high-level abstraction for building visualizations
to reduce software engineering know-how~\cite{heer2010declarative}.
Along these lines, based on the Grammar of Graphics~\cite{wilkinson2013grammar}, more recently developed visualization grammars such as Vega~\cite{satyanarayan2015reactive} and Vega-Lite~\cite{satyanarayan2016vega} support visualization design through declarative specifications, enabling rapid visualization design and prototyping.
NL4DV's primary goal is to return visualizations in response to NL queries.
To enable this, in addition to a structured representation of attributes and tasks inferred from a query, NL4DV also needs to return visualization specifications most relevant to an input query.
Given the conciseness of Vega-Lite and its growing usage in both web-based visualization systems and Python-based visual data analysis, NL4DV uses Vega-Lite as its underlying visualization grammar.

\subsection{Natural Language Processing Toolkits}

Toolkits like NLTK~\cite{loper2002nltk}, Stanford CoreNLP~\cite{manning2014stanford} and NER~\cite{finkel2005incorporating}, and spaCy~\cite{honnibal2017spacy} help developers perform NLP tasks such as part-of-speech (POS) tagging, entity recognition, and dependency parsing, among others.
However, since these are general-purpose toolkits, to implement visualization NLIs, developers need to learn to use the toolkit and also understand the underlying NLP techniques/concepts (e.g., knowing which dependency paths to traverse while parsing queries, understanding semantic similarity metrics).
Furthermore, to implement visualization systems, developers need to write additional code to convert the output from NLP toolkits into visualization-relevant concepts (e.g., attributes and values for applying data filters), which can be both complex and tedious.
Addressing these challenges, NL4DV internally uses NLTK~\cite{loper2002nltk}, Stanford CoreNLP~\cite{manning2014stanford}, and spaCy~\cite{honnibal2017spacy} but provides an API that encapsulates and hides the underlying NLP implementation details.
This allows visualization developers to focus more on front-end code pertaining to the user interface and interactions while invoking high-level functions to interpret NL queries.
\section{NL4DV Overview}

\begin{figure}[t!]
    \centering
    \includegraphics[width=\linewidth]{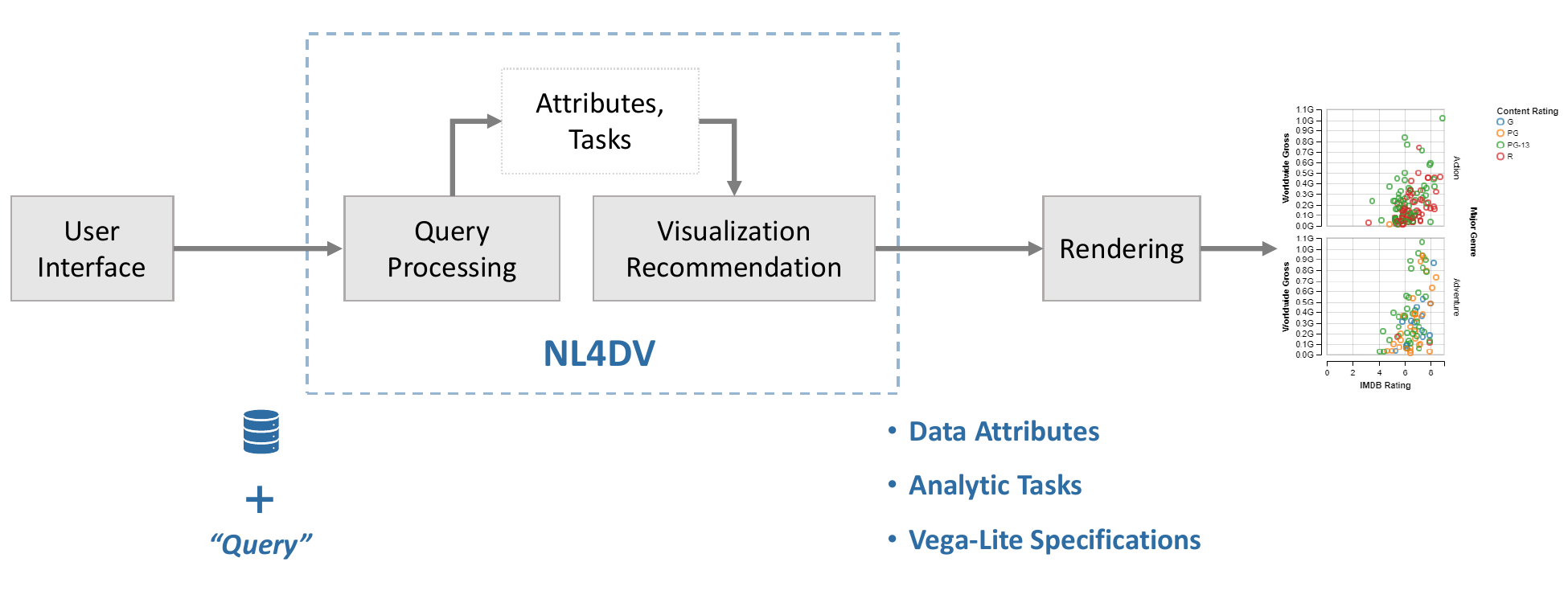}
    \vspace{-2em}
    \caption{\add{An overview of steps to generate visualizations based on NL queries.
    NL4DV encapsulates the query processing and visualization recommendation components, providing abstract functions to support their functionality.
    Once initialized with a dataset, NL4DV parses input NL queries and returns relevant information (in terms of data attributes and analytic tasks) and an ordered list of Vega-Lite specifications.
    }}
    \label{fig:overview}
    \vspace{-1.5em}
\end{figure}

Figure~\ref{fig:overview} presents an overview of a typical pipeline for implementing NLIs that generate visualizations in response to NL queries.
At a high-level, once an input query is collected through a \emph{User Interface}, a \emph{Query Processor} infers relevant information such as data attributes and analytic tasks from the input query.
This information is then passed to a \emph{Visualization Recommendation Engine} which generates a list of visualizations specifications relevant to the input query.
These specifications are finally rendered through a library (e.g., D3~\cite{bostock2011d3}) of the developer's choice.
\add{In the context of this pipeline, NL4DV provides a high-level API for processing NL queries and generating Vega-Lite specifications relevant to the input query.
Developers can choose to directly render the Vega-Lite specifications to create views (e.g., using Vega-Embed~\cite{vegaembed}) or use the attributes and tasks inferred by NL4DV to make custom changes to their system's interface.}

\subsection{Design Goals}

Four key design goals drove the development of NL4DV.
We compiled these goals based on a review of design goals and system implementations of prior visualization NLIs~\cite{sun2010articulate,gao2015datatone,setlur2016eviza,hoque2018applying,srinivasan2018orko,yu2019flowsense} and recent toolkits for supporting visualization development on new platforms and modalities (e.g,~\cite{schwab2019easypz,sicat2018dxr}).

\vspace{.5em}
\noindent{}\textbf{DG1.~Minimize NLP learning curve.}
NL4DV's primary target users are developers without a background or experience in working with NLP techniques.
Correspondingly, it was important to make the learning curve as flat as possible.
In other words, we wanted to enable developers to use the output of NL4DV without having to spend time learning about the mechanics of how information is extracted from NL queries.
In terms of toolkit design, this consideration translated to providing
high-level functions for interpreting NL queries and designing a response structure that was optimized for visualization system development by emphasizing visualization-related information such as analytic tasks (e.g., filter, correlation) and data attributes and values.

\vspace{.5em}
\noindent{}\textbf{DG2.~Generate modularized output and support integration with existing system components.}
By default, NL4DV recommends Vega-Lite specifications in response to NL queries.
However, a developer may prefer rendering visualizations using a different library such as D3 or may want to use a custom visualization recommendation engine (e.g.,~\cite{wongsuphasawat2016towards,moritz2018formalizing,lin2020dziban}), only leveraging NL4DV to identify attributes and/or tasks in the input query.
Supporting this goal required us to ensure that NL4DV's output was modularized (allowing developers to choose if they wanted attributes, tasks, and/or visualizations) and that developers do not have to significantly modify their existing system architecture to use NL4DV.
In terms of toolkit design, in addition to using a standardized grammar for visualizations (in our case, Vega-Lite), these considerations translated to devising a formalized representation of data attributes and analytic tasks that developers can programmatically parse to link NL4DV's output to other components in their system.

\vspace{.5em}
\noindent{}\textbf{DG3.~Highlight inference type and ambiguity.}
NL is often underspecified and ambiguous.
In other words, input queries may only include partial references to data attributes or may implicitly refer to intended tasks and visualizations (e.g., Figure~\ref{fig:teaser}b, c)~\cite{tory2019mean}.
Besides addressing these challenges from an interpretation standpoint, it was also important to make developers aware of the resulting uncertainty in NL4DV's output so they can choose to use/discard the output and provide appropriate visual cues (e.g., ambiguity widgets~\cite{gao2015datatone}) in their systems' interface.
In terms of toolkit design, this translated to structuring NL4DV's output so it
indicates whether information is inferred through an explicit (e.g., query substring matches an attribute name) or implicit (e.g., query refers to a data attribute through the attribute's values) reference, and highlights potential ambiguities in its response (e.g., two or more attributes map to the same word in an input query as in Figure~\ref{fig:teaser}c).

\vspace{.5em}
\noindent{}\textbf{DG4.~Support adding aliases and overriding toolkit defaults.}
Visualization systems are frequently used to analyze domain-specific datasets (e.g., sales, medical records, sports).
Given a domain, it is common for data attributes to have abbreviations or aliases (e.g., ``GDP" for \textit{Gross domestic product}, ``investment" for \textit{Capital}), or values that are unique to the dataset (e.g., the letter ``A" can refer to a value in a course grade dataset, but would be considered as a stopword and ignored
by most NLP algorithms by default).
In terms of toolkit design, these dataset-specific considerations translated to providing developers with helper functions to specify aliases or special word lists that NL4DV should consider/exclude for a given dataset.

\section{NL4DV Design and Implementation}


In this section, we detail NL4DV's design and implementation, highlighting key functions and describing how the toolkit interprets NL queries.
We defer discussing example applications developed using NL4DV to the following section.

Listing~\ref{listing:init-eg}
shows the basic Python code for using NL4DV.
Given a
\add{dataset (as a CSV, TSV, or JSON) and a}
query string, with a single function call \function{analyze\_query(\textit{query})}, NL4DV infers attributes, tasks, and visualizations, returning them as a JSON object (\textbf{DG1}).
Specifically,
NL4DV's response object has an 
\variable{attributeMap} composed of the inferred dataset attributes, a \variable{taskMap} composed of the inferred analytic tasks, and a \variable{visList}, a list of visualization specifications relevant to the input query.
By providing attributes, tasks, and visualizations as separate keys in the response object, NL4DV allows developers to selectively extract and use parts of its output (\textbf{DG2}).

\begin{listing}[t!]
\begin{minted}
[
baselinestretch=1,
fontsize=\scriptsize,
xleftmargin=15pt,
linenos,
breaklines
% ,style=tango
]
{python}
from nl4dv import NL4DV
nl4dv_instance = NL4DV(data_url="movies.csv")
response = nl4dv_instance.analyze_query("Show the relationship between budget and rating for Action and Adventure movies that grossed over 100M")
print(response)
\end{minted}
\vspace{-2em}
\begin{minted}[
style=tango
]
{json}
{
    "attributeMap": { ... },
    "taskMap": { ... },
    "visList": [ ... ]
}
\end{minted}
\vspace{-1.5em}
\caption{\textcolor{black}{Python code illustrating NL4DV's basic usage involving initializing NL4DV with a dataset (line 2) and analyzing a query string (line 3). The high-level structure of NL4DV's response is also shown.\vspace{-1.5em}}}
\label{listing:init-eg}
\end{listing}

\subsection{Data Interpretation}
\label{sec:data-interpretation}

Once initialized with a dataset (Listing~\ref{listing:init-eg}, line 2), NL4DV iterates through the underlying data item values to infer metadata including the attribute types ({\small{\faHashtag}}~\attrType{\textbf{Q}}uantitative, {\small{\faFont}}~\attrType{\textbf{N}}ominal, {\small{\faListUl}}~\attrType{\textbf{O}}rdinal, {\small{\faCalendar}}~\attrType{\textbf{T}}emporal) along with the range and domain of values for each attribute.
This attribute metadata is used when interpreting queries to infer appropriate analytic tasks and generate relevant visualization specifications.

Since NL4DV uses data values to infer attribute types, it may make erroneous interpretations.
For example, a dataset may have the attribute \emph{Day} with values in the range $[1, 31]$.
Detecting a range of integer values, by default, NL4DV will infer \emph{Day} as a quantitative attribute instead of temporal.
This misinterpretation can lead to NL4DV making poor design choices when selecting visualizations based on the inferred attributes (e.g., a quantitative attribute may result in a histogram instead of a line chart).
To overcome such issues caused by data quality or dataset semantics, NL4DV allows developers to verify the inferred metadata using \function{get\_metadata()}.
This function returns a hash map of attributes along with their inferred metadata.
If they notice errors, developers can use other helper functions (e.g., \function{set\_attribute\_type(\textit{attribute},\textit{type})}) to override the default interpretation (\textbf{DG4}).


\subsection{Query Interpretation}

\begin{figure}[b!]
    \centering
    \includegraphics[width=\linewidth]{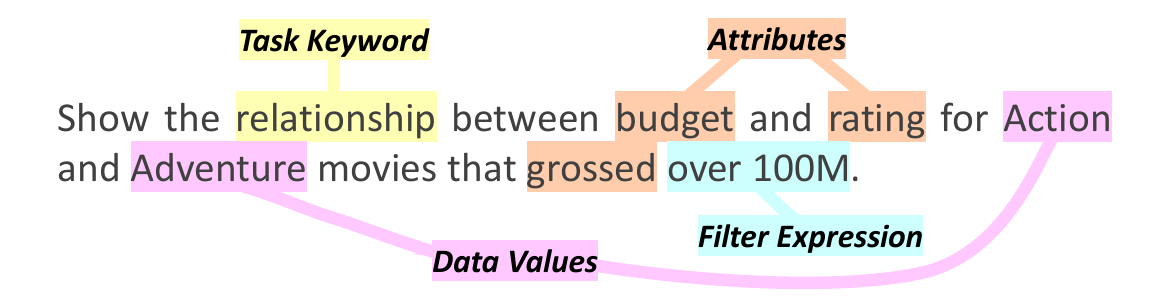}
    \caption{An illustration of query phrases that NL4DV identifies while interpreting NL queries.}
    \label{fig:parse-eg}
\end{figure}

\begin{figure}[t!]
    \centering
    \vspace{-1.5em}
    \includegraphics[width=\linewidth]{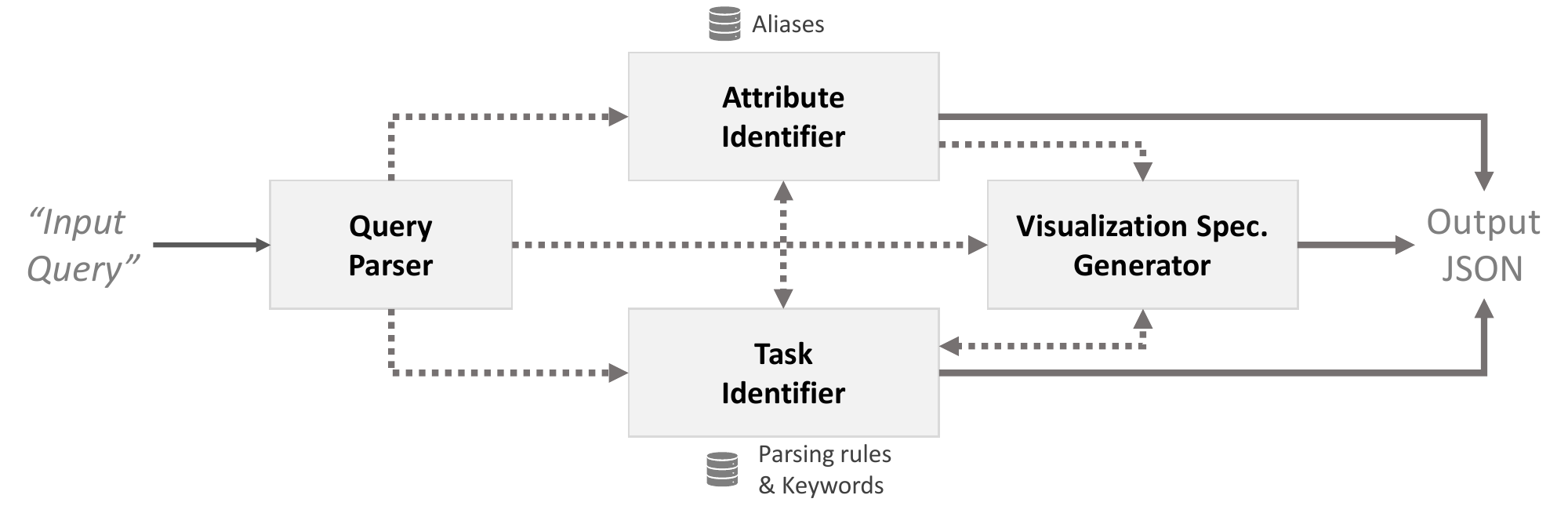}
    \caption{NL4DV's architecture. The arrows indicate the flow of information between different modules.}
    \label{fig:nl4dv-architecture}
    \vspace{-1.5em}
\end{figure}

To generate visualizations in response to a query, visualization NLIs need to identify informative phrases in the query that map to relevant concepts like data attributes and values, analytic tasks, and visualization types, among others.
Figure~\ref{fig:parse-eg} shows the query in Listing~\ref{listing:init-eg}
\canremove{``\textit{Show the relationship between budget and rating for Action and Adventure movies that grossed over 100M}"}
with such annotated phrases (we use this query as a running example throughout this section to describe NL4DV's query interpretation strategy).
To identify relevant phrases and generate the \variable{attributeMap}, \variable{taskMap}, and \variable{visList}, NL4DV performs four steps: 1) \textit{query parsing}, 2) \textit{attribute inference}, 3) \textit{task inference}, and 4) \textit{visualization specification generation}.
Figure~\ref{fig:nl4dv-architecture} gives an overview of NL4DV's underlying architecture.
Below we describe the individual query interpretation steps (task inference is split into two steps to aid explanation) and summarize the pipeline in Figure~\ref{fig:main-eg}.

\subsubsection{Query Parsing}

\begin{figure*}[t!]
    \centering
    \includegraphics[width=.9\linewidth]{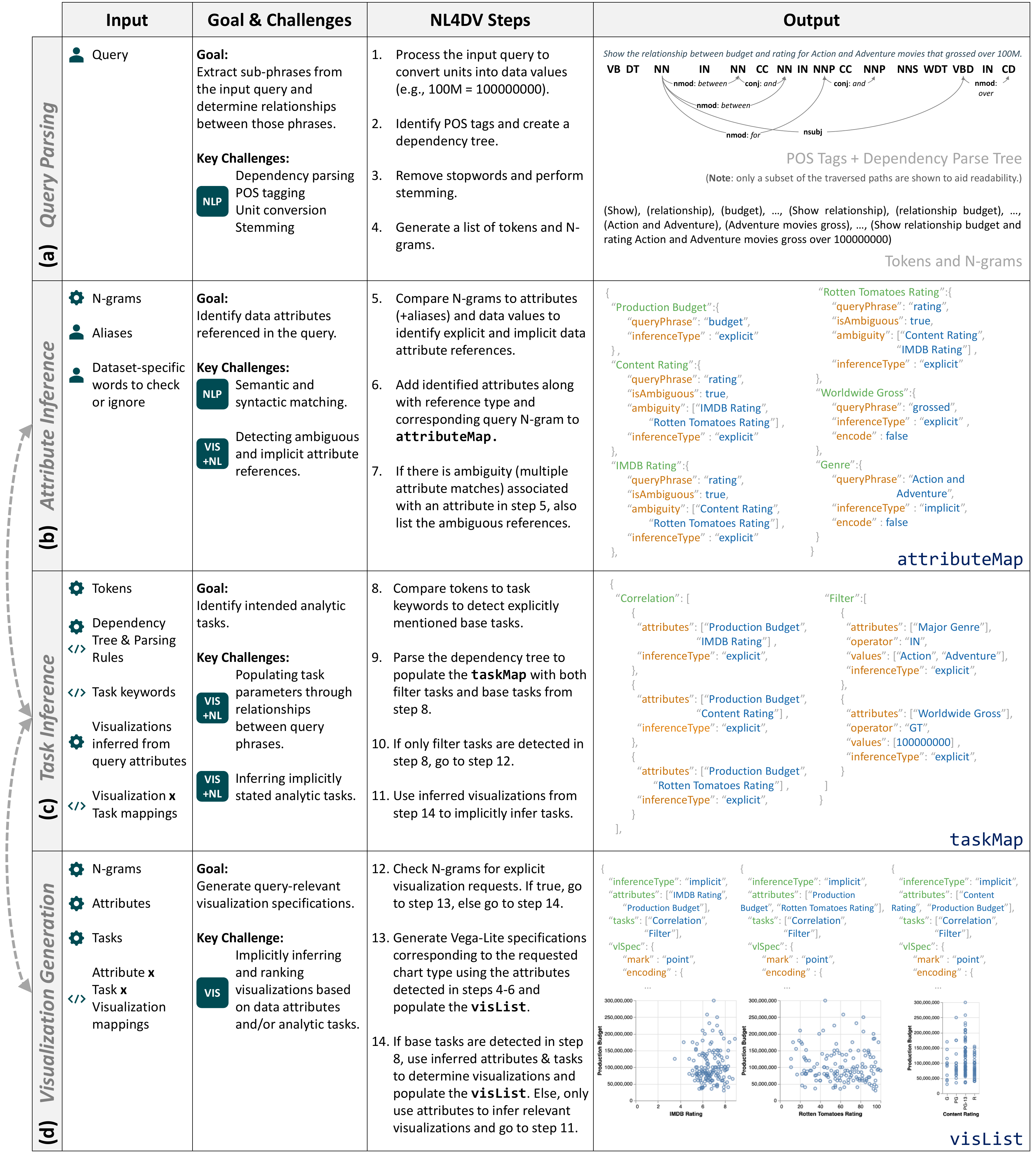}
    \vspace{-.5em}
    \caption{
    Summary of the query interpretation pipeline triggered by \function{analyze\_query(\textit{query})}.
    Information flows sequentially across stages unless explicitly indicated by bi-directional arrows.
    Input to different stages is {{\textcolor{darkteal}{\faUser}}} provided externally by developers, {{\textcolor{darkteal}{\faCog}}} generated by other stages of NL4DV's pipleline, or  {{\textcolor{darkteal}{\faCode}}} preconfigured into NL4DV.
    The figure also highlights the key goal and implementation challenges for each stage (\implementationchallenge{NLP}: general NLP challenge, \implementationchallenge{VIS+NL}: challenge specific to visualization NLIs, \implementationchallenge{VIS}: visualization design/recommendation challenge).
    NL4DV internally tackles these challenges, providing visualization developers with a high-level API for query interpretation.
    }
    \label{fig:main-eg}
    \vspace{-1em}
\end{figure*}

The query parser runs a series of NLP functions on the input string to extract details that can be used to detect relevant phrases.
In this step, NL4DV first preprocesses the query to convert any special symbols or characters into dataset-relevant values (e.g., converting \textit{100M} to the number \textit{100000000}).
Next, the toolkit identifies the POS tags for each token (e.g., \textit{NN}: Noun, \textit{JJ}: Adjective, \textit{CC}: Coordinating Conjunction) using Stanford's CoreNLP~\cite{manning2014stanford}.
Furthermore, to understand the relationship between different phrases in the query, NL4DV uses CoreNLP's dependency parser to create a dependency tree.
Then, with the exception of conjunctive/disjunctive terms (e.g., `and', `or') and some prepositions (e.g., `between', `over') and adverbs (e.g., `except', `not'), NL4DV trims the input query by removing all stop words and performs stemming (e.g., `\textit{grossed}' $\rightarrow$ `\textit{gross}').
Lastly, the toolkit generates all \emph{N-grams}
from the trimmed query string.
The output from the query parser (POS tags, dependency tree, N-grams) is shown in Figure~\ref{fig:main-eg}a and is used internally by NL4DV during the remaining stages of query interpretation.



\subsubsection{Attribute Inference}

After parsing the input query, NL4DV looks for data attributes that are mentioned both \emph{explicitly} (e.g., through direct references to attribute names) and \emph{implicitly} (e.g., through references to an attribute's values).
Developers can also configure aliases (e.g., `Investment' for \textit{Production Budget}) to support dataset- and domain-specific attribute references (\textbf{DG4}).
To do so, developers can provide a JSON object consisting of attributes (as keys) and lists of aliases (as values).
This object can be passed through the optional parameters \textbf{alias\_map} or \textbf{alias\_map\_url} when initializing NL4DV (Listing~\ref{listing:init-eg}, line 2) or using the helper function \function{set\_alias\_map(\textit{alias\_map}, \textit{url}="")}.

To infer attributes, NL4DV iterates through the N-grams generated by the query parser, checking for both syntactic (e.g., misspelled words) and semantic (e.g., synonyms) similarity between N-grams and a lexicon composed of data attributes, aliases, and values.
To check for syntactic similarity, NL4DV computes the cosine similarity $Sim\textsubscript{cos}(i, j)$ between a N-gram \emph{i} and a tokenized lexical entity \emph{j}.
The possible values for $Sim\textsubscript{cos}(i, j)$ range from [0,1] with 1 indicating that strings are equivalent.
For semantic similarity, the toolkit checks for the Wu-Palmer similarity score~\cite{wu1994verbs} $Sim\textsubscript{wup}(i, j)$ between a N-gram \emph{i} and a tokenized lexical entry \emph{j}.
This score returns the distance between stemmed versions of \emph{p} and \emph{a} in the WordNet
graph~\cite{miller1995wordnet}, and is a value in the range $(0,1]$, with higher values implying greater similarity.
If $Sim\textsubscript{cos}(i, j)$ or $Sim\textsubscript{wup}(i, j) \geq 0.8$, NL4DV maps the N-gram \emph{i} to the attribute corresponding to \emph{j}, also adding the attribute as a key in the \variable{attributeMap}.

As shown in Figure~\ref{fig:main-eg}b, the \variable{attributeMap} is structured such that besides the attributes themselves, for each attribute, developers can also identify: (1) query substrings that led to an attribute being detected (\sKey{queryPhrase}) along with (2) the type of reference (\sKey{inferenceType}), and (3) ambiguous matches (\sKey{ambiguity}).
For instance, given the query in Figure~\ref{fig:parse-eg}, NL4DV detects the attributes \key{Production Budget} (based on `\val{budget}'), \key{Content Rating}, \key{IMDB Rating}, \key{Rotten Tomatoes Rating} (\sKey{ambiguity} caused by the word `\val{rating}'), \key{Worldwide Gross} (based on `\val{grossed}'), and \key{Genre} (based on the values `\val{Action}' and `\val{Adventure}').
Furthermore, since \textit{Genre} is referenced by its values, it is marked as \val{implicit} whereas the other attributes are marked as \val{explicit} (\textbf{DG3}).

\subsubsection{Explicit Task Inference}

\add{After detecting N-grams mapping to data attributes, NL4DV checks the remaining N-grams for references to analytic tasks.}
NL4DV currently identifies five low-level analytic tasks~\cite{amar2005low} including four base tasks: \emph{Correlation}, \emph{Distribution}, \emph{Derived Value}, \emph{Trend}, and a fifth \emph{Filter} task.
We separate base tasks from filter since base tasks are used to determine appropriate visualizations (e.g., correlation maps to a scatterplot) whereas filters are applied across different types of visualizations.
We focus on these five tasks as a starting set since they are commonly supported in prior NLIs~\cite{sun2010articulate,gao2015datatone,setlur2019inferencing,yu2019flowsense,kassel2018valletto} and are most relevant to NL4DV's primary goal of supporting visualization specification through NL (as opposed to interacting with a given chart~\cite{setlur2016eviza,srinivasan2018orko} or question answering~\cite{kim2020answering}).

While filters may be detected via data values (e.g.,~`Action', `Comedy'), to detect base tasks, NL4DV compares the query tokens to a predefined list of task keywords (e.g., `correlate', `relationship', etc., for the \textit{Correlation} task, `range', `spread', etc., for the \textit{Distribution} task, `average', `sum', etc., for \textit{Derived Value}).
Merely detecting references to attributes, values, and tasks is insufficient to infer user intent, however.
To model relationships between query phrases and populate task details, NL4DV leverages the POS tags and the dependency tree generated by the query parser.
Specifically, using the token type and dependency type (e.g., \texttt{nmod}, \texttt{conj}, \texttt{nsubj}) and distance, NL4DV identifies mappings between attributes, values, and tasks.
These mappings are then used to model the \variable{taskMap}.
\add{The task keywords and dependency parsing rules were defined based on the query patterns and examples from prior visualization NLIs~\cite{sun2010articulate,setlur2016eviza,gao2015datatone,hoque2018applying,yu2019flowsense} as well as $\sim$200 questions collected by Amar et al.~\cite{amar2005low} when formulating their analytic task taxonomy.}

The \variable{taskMap} contains analytic tasks as keys.
Tasks are broken down as a list of objects that include an \ssKey{inferenceType} field to indicate if a task was stated explicitly (e.g., through keywords) or derived implicitly (e.g., if a query requests for a line chart, a \textit{trend} task may be implied) and parameters to apply when executing a task.
These include the \ssKey{attributes} a task maps to, the \ssKey{operator} to be used (e.g., \val{GT}, \val{EQ}, \val{AVG}, \val{SUM}), and \ssKey{values}.
If there are ambiguities in task parameters (e.g., the word `fiction' may refer to the values `Science Fiction,' `Contemporary Fiction,' `Historical Fiction'), NL4DV adds additional fields (e.g., \ssKey{isValueAmbiguous}$=$\val{true}) to highlight them (\textbf{DG3}).
In addition to the tasks themselves, this structuring of the \variable{taskMap} allows developers to detect: (1) the parameters needed to execute a task (\ssKey{attributes}, \ssKey{operator}, \ssKey{values}), (2) operator- and value-level ambiguities (e.g., \ssKey{isValueAmbiguous}), and (3) if the task was stated explicitly or implicitly (\ssKey{inferenceType}).

Consider the \variable{taskMap} (Figure~\ref{fig:main-eg}c) for the query in Figure~\ref{fig:parse-eg}.
Using the dependency tree in Figure~\ref{fig:main-eg}a, NL4DV infers that the word `relationship' maps to the \key{Correlation} task and links to the tokens `budget' and `rating' which are in-turn linked by the conjunction term `and.'
Next, referring back to the \variable{attributeMap}, NL4DV maps the words `budget' and `rating' to their respective data attributes, adding three objects corresponding to correlations between the \ssKey{attributes} [\val{Production Budget, IMDB Rating}], [\val{Production Budget, Content Rating}], and [\val{Production Budget, Rotten Tomatoes Rating}] to the \textit{correlation} task.
Leveraging the tokens `Action' and `Adventure', NL4DV also infers that the query refers to a \key{Filter} task on the \ssKey{attribute} \val{Genre}, where the \ssKey{values} are \textit{in the list} (\val{IN}) [\val{Action}, \val{Adventure}]. 
Lastly, using the dependencies between tokens in the phrase `\textit{gross over 100M},' NL4DV adds an object with the \ssKey{attribute} \val{Worldwide Gross}, the \textit{greater than} (\val{GT}) \ssKey{operator}, and \val{100000000} in the \ssKey{values} field.
While populating \textit{filter} tasks, NL4DV also updates the corresponding attributes in the \variable{attributeMap} with the key \sKey{encode}$=$\val{False} (Figure~\ref{fig:main-eg}b).
This helps developers detect that an attribute is used for filtering and is not visually encoded in the recommended charts.

\subsubsection{Visualization Generation}
\label{sec:vis-spec-gen}

NL4DV uses Vega-Lite as the underlying visualization grammar.
The toolkit currently supports the Vega-Lite \sKey{marks}: \val{bar}, \val{tick}, \val{line}, \val{area}, \val{point}, \val{arc}, \val{boxplot}, \val{text} and \sKey{encodings}: \val{x}, \val{y}, \val{color}, \val{size}, \val{column}, \val{row}, \val{theta} to visualize up to three attributes at a time.
This combination of marks and encodings allows NL4DV to support a range of common visualization types including \emph{histograms}, \emph{strip plots}, \emph{bar charts} (including stacked and grouped bar charts), \emph{line} and \emph{area charts}, \emph{pie charts}, \emph{scatterplots}, \emph{box plots}, and \emph{heatmaps}.
To determine visualizations relevant to the input query, NL4DV checks the query for explicit requests for visualization types (e.g., Figure~\ref{fig:teaser}a) or implicitly infers visualizations from attributes and tasks (e.g., Figures~\ref{fig:teaser}b,~\ref{fig:teaser}c, and~\ref{fig:parse-eg}).

Explicit visualization requests are identified by comparing query N-grams to a predefined list of visualization keywords (e.g., `scatterplot', `histogram', `bar chart').
For instance, the query in Figure~\ref{fig:teaser}a specifies the visualization type through the token `histogram,' leading to NL4DV setting \val{bar} as the \sKey{mark} type and binned \val{IMDB Rating} as the \val{\texttt{x}} \sKey{encoding} in the underlying Vega-Lite specification.

To implicitly determine visualizations, NL4DV uses a combination of the attributes and tasks inferred from the query.
NL4DV starts by listing all possible visualizations using the detected attributes by applying well-known mappings between attributes and visualizations (Table~\ref{tab:tasks}).
\add{These mappings are preconfigured within NL4DV based on heuristics used in prior systems like Show Me~\cite{mackinlay2007show} and Voyager~\cite{wongsuphasawat2015voyager,wongsuphasawat2017voyager}.}
As stated earlier, when generating visualizations from attributes, NL4DV does not visually encode the attributes used as filters.
Instead, filter attributes are added as a \val{\texttt{filter}} \sKey{transform} in Vega-Lite.
Doing so helps avoid a combinatorial explosion of attributes when a query includes multiple filters (e.g., including the filter attributes for the query in Figure~\ref{fig:parse-eg} would require generating visualizations that encode four attributes instead of two).

Besides attributes, if tasks are explicitly stated in the query, NL4DV uses them as an additional metric to modify, prune, and/or rank the generated visualizations.
Consider the query in Figure~\ref{fig:parse-eg}.
Similar to the query in Figure~\ref{fig:teaser}c, if only attributes were used to determine the charts, NL4DV would output two scatterplots (for \attrType{Q}x\attrType{Q}) and one bar chart (for \attrType{N}x\attrType{Q}).
However, since the query contains the token `relationship,' which maps to a \textit{Correlation} task, NL4DV enforces a scatterplot as the chart type, setting the \sKey{mark} in the Vega-Lite specifications to \val{point}.
\add{Furthermore, because correlations are more apparent in \attrType{Q}x\attrType{Q} charts, NL4DV also ranks the two \attrType{Q}x\attrType{Q} charts higher, returning the three visualization specifications shown in Figure~\ref{fig:main-eg}d.
These Task~x~Visualization mappings (Table~\ref{tab:tasks}) are
configured within NL4DV based on prior visualization systems~\cite{moritz2018formalizing,casner1991task,gotz2009behavior} and studies~\cite{kim2018assessing,saket2018task}.}

\begin{table}[t!]
\centering
\resizebox{\linewidth}{!}{%
\begin{tabular}{lll}
\hline
\multirow{2}{*}{\textbf{\begin{tabular}[c]{@{}l@{}}Attributes\\ (\texttt{x}, \texttt{y}, \texttt{color}/\texttt{size}/\texttt{row}/\texttt{column})\end{tabular}}} &
  \multirow{2}{*}{\textbf{Visualizations}} &
  \multirow{2}{*}{\textbf{Task}} \\
                                        &                                         &                                \\ \hline
\multirow{2}{*}{Q x Q x \{N, O, Q, T\}} & \multirow{2}{*}{Scatterplot}            & \multirow{2}{*}{Correlation}   \\
                                        &                                         &                                \\
\multirow{2}{*}{N, O x Q x \{N, O, Q, T\}} & \multirow{2}{*}{Bar Chart}              & \multirow{2}{*}{Derived Value} \\
                                        &                                         &                                \\
Q, N, O x \{N, O, Q, T\} x \{Q\} &
  \begin{tabular}[c]{@{}l@{}}Strip Plot, Histogram,\\Bar Chart, Heatmap\end{tabular} &
  Distribution \\
\multirow{2}{*}{T x \{Q\} x \{N, O\}}   & \multirow{2}{*}{Line Chart} & \multirow{2}{*}{Trend}         \\
                                        &                                         &                                \\ \hline
\end{tabular}%
}
\vspace{-.5em}
\caption{Attribute (+encodings), visualization, and task mappings preconfigured in NL4DV.
Attributes in curly brackets \{are optional\}.
Note that these defaults can be overridden via explicit queries. For instance, ``\textit{Show average gross across
genres as a scatterplot}" will create a scatterplot instead of a bar chart with \textit{Genre} on the \texttt{x}- and AVG(\textit{Worldwide Gross}) on the \texttt{y}-axis.
For unsupported attribute combinations and tasks, NL4DV resorts to a table-like view created using Vega-Lite's \textit{text} \texttt{mark}.
}
\label{tab:tasks}
\vspace{-1.5em}
\end{table}

NL4DV complies the inferred visualizations into a \variable{visList} (Figure~\ref{fig:main-eg}d).
Each object in this list is composed of a \key{vlSpec} containing the Vega-Lite specification for a chart, an \key{inferenceType} field to highlight if a visualization was requested explicitly or implicitly inferred by NL4DV, and a list of \key{attributes} and \key{tasks} that a visualization maps to.
Developers can use the \variable{visList} to directly render visualizations in their systems (via the \key{vlSpec}).
\canremove{Alternatively, ignoring the \variable{visList}, developers can also extract only attributes and tasks using the \variable{attributeMap} and \variable{taskMap}, and feed them as input to other visualization recommendation engines (e.g.,~\cite{wongsuphasawat2016towards,lin2020dziban}) (\textbf{DG2}).}

\subsubsection{Implicit Task Inference}
When the input query lacks explicit keywords referring to analytic tasks, NL4DV first checks if the query requests for a specific visualization type.
If so, the toolkit uses mappings between Visualizations~x~Tasks in Table~\ref{tab:tasks} to infer tasks (e.g., \textit{distribution} for a histogram, \textit{trend} for a line chart, \textit{correlation} for a scatterplot).

Alternatively, if the query only mentions attributes, NL4DV first lists possible visualizations based on those attributes.
Then, using the inferred visualizations, the toolkit implicitly infers tasks (again leveraging the Visualization~x~Task mappings in Table~\ref{tab:tasks}).
Consider the example in Figure~\ref{fig:teaser}c.
In this case, the tasks \textit{Correlation} and \textit{Derived Value} are inferred based on the two scatterplots and one bar chart generated using the attribute combinations \attrType{Q}{\small{x}}\attrType{Q} and \attrType{N}{\small{x}}\attrType{Q}, respectively.
In such cases where the tasks are implicitly inferred through visualizations, NL4DV also sets their \ssKey{inferenceType} in the \variable{taskMap} to \val{implicit}.




\section{Example Applications}



\subsection{Using NL4DV in Jupyter Notebook}

\begin{figure}[t!]
    \centering
    \includegraphics[width=\linewidth]{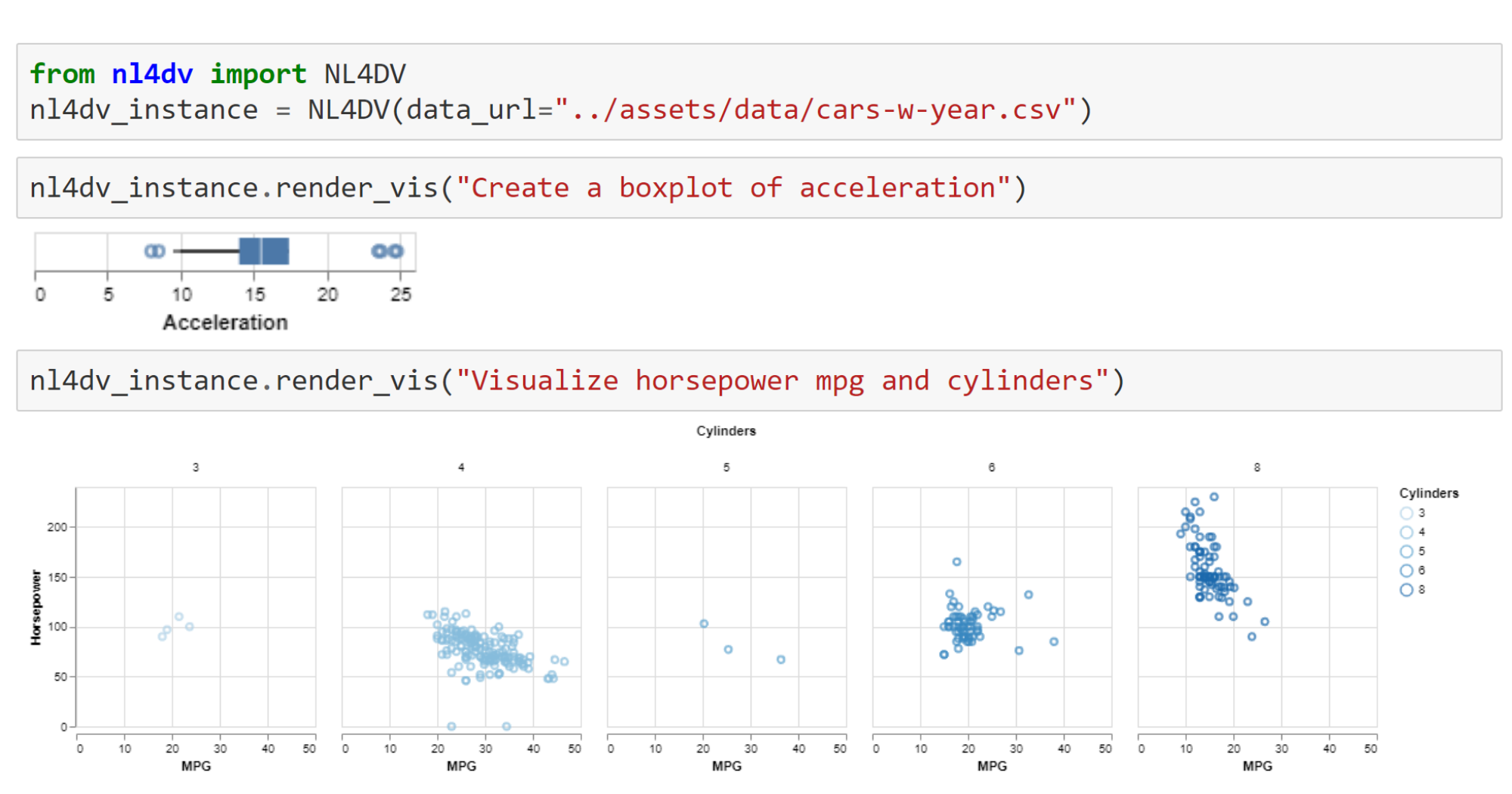}
    \vspace{-2em}
    \caption{NL4DV being used to specify visualizations through NL in Python within a Jupyter Notebook.}
    \label{fig:example-notebook}
    \vspace{-1em}
\end{figure}

Since NL4DV generates Vega-Lite specifications, in environments that support rendering Vega-Lite charts, the toolkit can be used to create visualizations through NL in Python.
Specifically, NL4DV provides a wrapper function \function{render\_vis(\textit{query})} that automatically renders the first visualization in the \variable{visList}.
By rendering visualizations in response to NL queries in environments like Jupyter Notebook, NL4DV enables novice Python data scientists and programmers to conduct visual analysis without needing to learn about visualization design or Python visualization packages (e.g., Matplotlib, Plotly).
Figure~\ref{fig:example-notebook} shows an instance of a Jupyter Notebook demonstrating the use of NL4DV to create visualizations for a cars dataset.
For the first query ``\textit{Create a boxplot of acceleration}," detecting an explicit visualization request, NL4DV renders a \textit{box plot} showing values for the attribute \textit{Acceleration}.
For the second query ``\textit{Visualize horsepower mpg and cylinders}", NL4DV implicitly selects a scatterplot as the appropriate visualization using the inferred attributes ({\small{\faHashtag}} \textit{Horsepower}, {\small{\faHashtag}} \textit{MPG}, {\small{\faListOl}} \textit{Cylinders}).

\begin{figure}[t!]
    \centering
    \includegraphics[width=\linewidth]{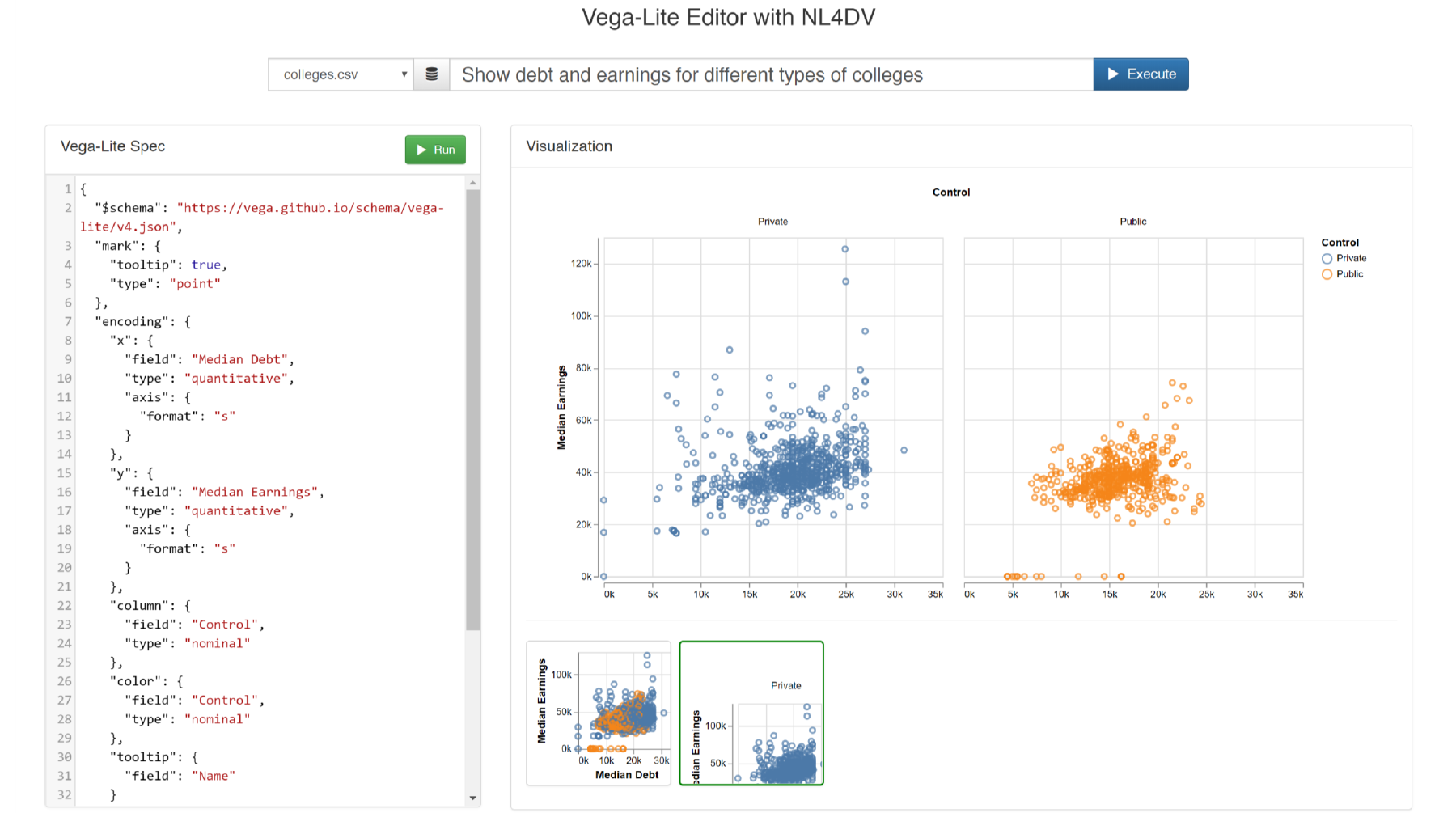}
    \vspace{-1.5em}
    \caption{A Vega-Lite editor that supports NL-based chart specification and presents design alternatives using the \variable{visList} returned by NL4DV.
    Here, the query ``\textit{Show debt and earnings for different types of colleges}" issued in the context of a U.S.~colleges dataset results in the system suggesting a colored scatterplot and a colored~+~faceted scatterplot.
    The faceted scatterplot is selected as the active chart by the user.
    }
    \label{fig:example-vl-editor}
    \vspace{-.75em}
\end{figure}

\subsection{Creating Visualization Systems with NL4DV}

The above example illustrates how the \variable{visList} generated by NL4DV is used to create visualizations in Python-based data science environments.
The following two examples showcase how NL4DV can assist the development of web-based NLIs for data visualization.

\subsubsection{NL-Driven Vega-Lite Editor}

Although the declarative nature of Vega-Lite makes it an intuitive way to specify visualizations, novices unaware of visualization terminology may need to spend a significant amount of time and effort to look at examples and learn the specification grammar.
NL input presents itself as a promising solution in such scenarios to help onboard users for learning Vega-Lite.
With a NLI, users can load their data and express their intended chart through NL.
The system in response can present both the chart and the corresponding Vega-Lite specification, allowing users to learn the underlying grammar through charts of their interest.
Figure~\ref{fig:example-vl-editor} illustrates this idea implemented as an alternative version of the Vega-Lite editor~\cite{vegaeditor}, supporting NL input.
Users can enter queries through the text input box at the top of the page to specify charts and edit the specification on the left to modify the resulting visualization.
Besides the main visualization returned in response to an NL query, the interface also presents a alternative visualizations using different encodings similar to the Voyager systems~\cite{wongsuphasawat2015voyager,wongsuphasawat2017voyager}.

This example is developed following a classic client-server architecture and is implemented as a Python Flask~\cite{flask} application.
From a development standpoint, the client-side of this application is written from scratch using HTML and JavaScript.
On the Python server-side, a single call is made to NL4DV's \function{analyze\_query(\textit{query})} function where the \texttt{\textit{query}} is collected and passed via JavaScript.
As shown earlier (Listing~\ref{listing:init-eg}), this function returns a JSON object composed of the \variable{attributeMap}, \variable{taskMap}, and \variable{visList}.
For this example, the \variable{visList} returned by NL4DV is parsed in JavaScript to render the main chart along with the alternative designs (Listing~\ref{listing:vl-editor-js}, lines 5-8).
The visualizations are rendered using Vega-Embed~\cite{vegaembed}.

\begin{listing}[t!]
\begin{minted}
[
baselinestretch=1,
fontsize=\scriptsize,
xleftmargin=15pt,
% xleftmargin=2.5pt,
linenos,
breaklines,
style=manni
]
{javascript}

$.post("/analyzeQuery", {"query": query})
  .done(function (responseString) {
    let nl4dvResponse = JSON.parse(responseString);
    let visList = nl4dvResponse['visList'];
    // render visList[0]['vlSpec'] as default chart
    for(let visObj of visList){
      // add visObj['vlSpec'] as a thumbnail in the bottom panel displaying all possible designs
    }
  });
  
\end{minted}
\vspace{-1.5em}
\caption{\textcolor{black}{JavaScript code to parse NL4DV's output to create the Vega-Lite editor application shown in Figure~\ref{fig:example-vl-editor}.}
\vspace{-.5em}}
\label{listing:vl-editor-js}
\end{listing}

\subsubsection{Recreating Ambiguity Widgets in DataTone}

Consider a second example where we use NL4DV to replicate features of the DataTone system~\cite{gao2015datatone}.
Given a NL query, DataTone identifies ambiguities in the query and surfaces them via ``ambiguity widgets" (dropdown menus) that users can interact with to clarify their intent.

\begin{figure}[t!]
    \centering
    \includegraphics[width=\linewidth]{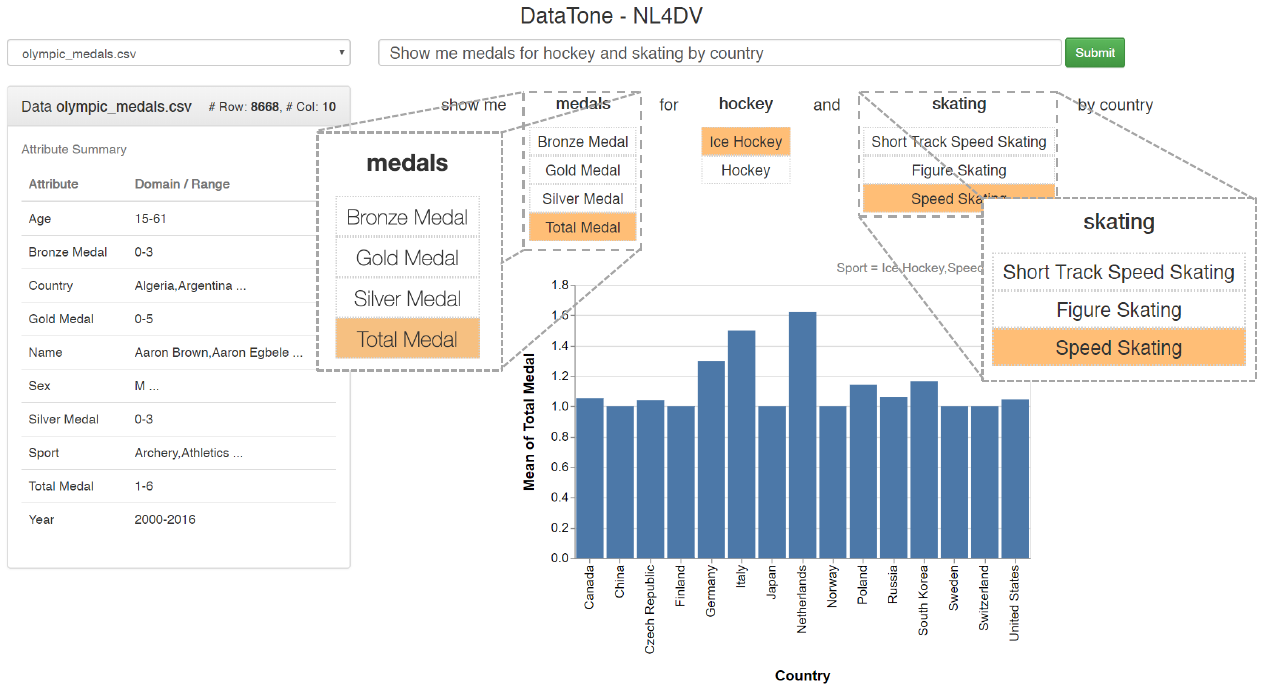}
    \caption{A sample interface illustrating how NL4DV can be used to replicate DataTone's~\cite{gao2015datatone} ambiguity widgets.
    }
    \label{fig:example-datatone}
    \vspace{-2em}
\end{figure}

Figure~\ref{fig:example-datatone} shows a DataTone-like interface implemented using NL4DV.
This system is also implemented as a Flask web-application using HTML and JavaSript on the client-side.
The example in Figure~\ref{fig:example-datatone} illustrates the result of executing the query ``\textit{Show me medals for hockey and skating by country}" against an Olympics medal winners dataset (query reused from the DataTone paper~\cite{gao2015datatone}).
Here, `medals' is an ambiguous reference to four data attributes---the three medal types (\textit{Bronze}, \textit{Silver}, \textit{Gold}) and the \textit{Total Medals}.
Similarly, `hockey' and `skating' are value-level ambiguities corresponding to the \textit{Sport} attribute (e.g., `hockey'$=$[\textit{Ice Hockey}, \textit{Hockey}]).

Similar to the Vega-Lite editor application, the server-side code only involves initializing NL4DV with the active dataset and making a call to the \function{analyze\_query(\textit{query})} function to process user queries.
As detailed in Listing~\ref{listing:datatone-js-parsing}, on the client-side, to highlight attribute- and value-level ambiguities in the query, we parse the \variable{attributeMap} and \variable{taskMap} returned by NL4DV in JavaScript, checking the \texttt{isAmbiguous} fields.
Vega-Embed is once again used to render the \texttt{vlSpec}s returned as part of NL4DV's \variable{visList}.
Note that we only focus on data ambiguity widgets in this example, not displaying design ambiguity widgets (e.g., dropdown menus for switching between visualization types).
To generate design ambiguity widgets, however, developers can parse the \variable{visList}, converting the Vega-Lite \texttt{marks} and \texttt{encodings} into dropdown menu options.


\begin{listing}[t!]
\begin{minted}
[
baselinestretch=1,
fontsize=\scriptsize,
xleftmargin=15pt,
% xleftmargin=2.5pt,
linenos,
breaklines,
style=manni
]
{javascript}

$.post("/analyzeQuery", {"query": query})
  .done(function (responseString) {
    let nl4dvResponse = JSON.parse(responseString);
    let attributeMap = nl4dvResponse['attributeMap'],
    taskMap = nl4dvResponse['taskMap'];
    for(let attr in attributeMap){
      if(attributeMap[attr]['isAmbiguous']){
        // add attr and attributeMap[attr]['ambiguity'] to attribute-level ambiguity widget corresponding to the attributeMap[attr]['queryPhrase']
      }
    }
    ...
  });
  
\end{minted}
\vspace{-1.75em}
\caption{\textcolor{black}{JavaScript code to parse NL4DV's output and generate attribute-level ambiguity widgets (highlighted in Figure~\ref{fig:example-datatone}-left).
A similar logic is used to iterate over the \variable{taskMap} when creating value-level ambiguity widgets (highlighted in Figure~\ref{fig:example-datatone}-right) for filtering.\vspace{-2.75em}
}}
\label{listing:datatone-js-parsing}
\end{listing}

\subsection{Adding NL Input to an Existing Visualization System}

\begin{figure}[t!]
    \centering
    \vspace{-1em}
    \includegraphics[width=.59\linewidth]{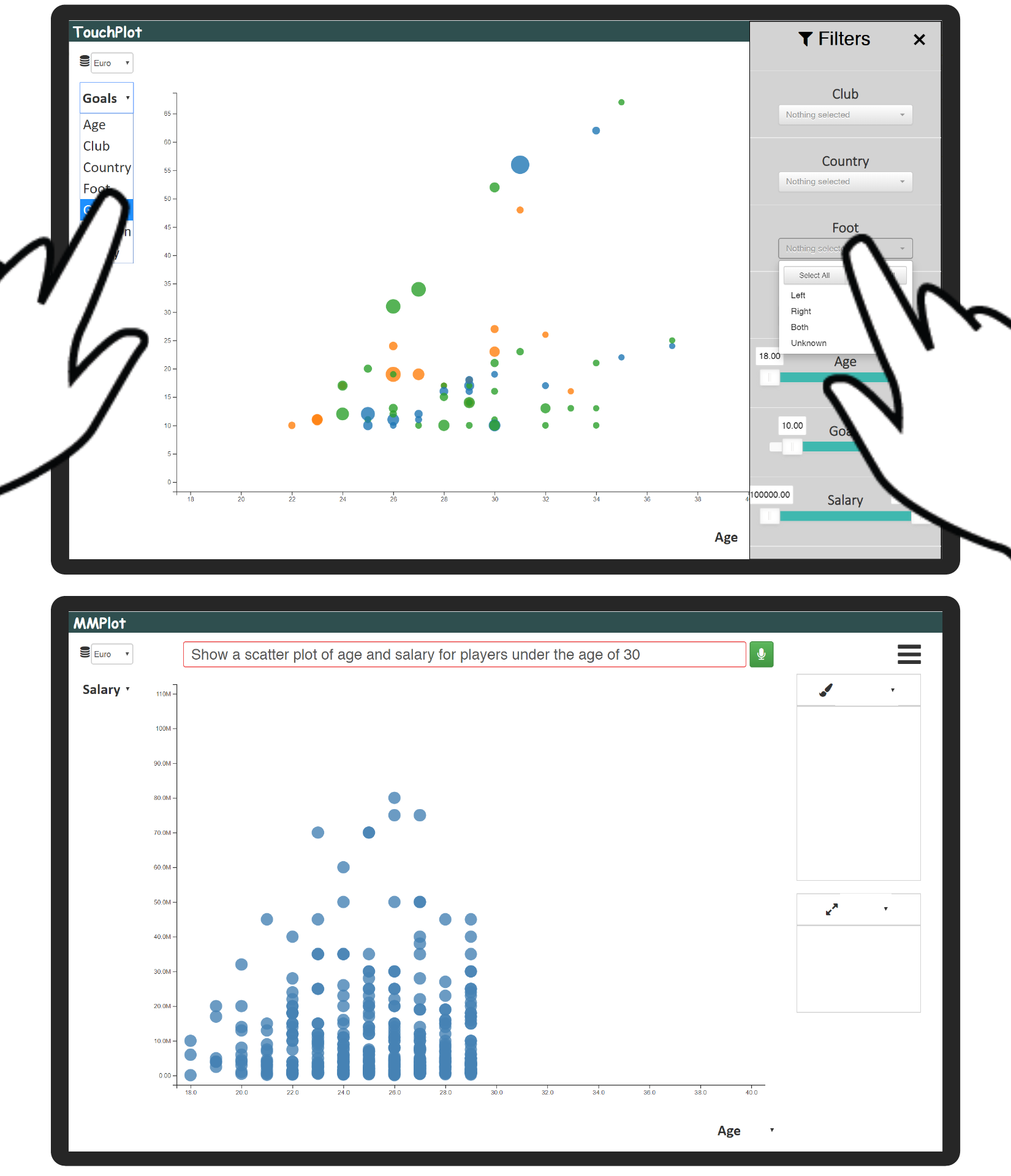}
    \caption{(Top) \textsc{TouchPlot} interface supporting interaction through touch and control panels.
    (Bottom) \textsc{MMPlot} interface supporting multimodal interactions.
    Here, the user has specified a new scatterplot and applied a filter through a single query ``\textit{Show a scatter plot of age and salary for players under the age of 30}."
    }
    \label{fig:example-mmplot}
    \vspace{-1em}
\end{figure}

NL4DV can also be used to augment existing visualization systems with NL input.
As an example, consider \textsc{TouchPlot} (Figure~\ref{fig:example-mmplot}-top), a touch-based scatterplot visualization system running on a tablet.
We modeled \textsc{TouchPlot} after the interface and capabilities of the scatterplot visualization system, Tangere~\cite{sadana2014designing}.
Specifically, users can select points and zoom/pan by interacting directly with the chart canvas, bind attributes to the position, color, and size encodings using dropdown menus on axes and legends, or apply filters using a side panel.
\textsc{TouchPlot} is implemented using HTML and JavaScript, and D3 is used for creating the visualization.

Recent work has shown that complementing touch interactions with speech can support a more fluid interaction experience during visual analysis on tablets~\cite{srinivasan2020inchorus}.
For example, while touch can support fine-grained interactions with marks, speech can allow specifying filters without having to open and interact with the side panel, saving screen space and preserving the user workflow.
To explore such fluid interactions, we developed \textsc{MMPlot} (Figure~\ref{fig:example-mmplot}-bottom), a modified version of \textsc{TouchPlot} that supports multimodal touch and speech input.
In addition to touch interactions, \textsc{MMPlot} allows issuing speech commands to specify charts (e.g., ``\textit{Correlate age and salary by country}") and filter points (e.g., ``\textit{Remove players over the age of 30}").

To support these interactions, we record speech input and convert it to a text string using the Web Speech API~\cite{webspeechapi}.
This query string is then passed to the server, where we make a call to NL4DV's \function{analyze\_query(\textit{query})}. 
By parsing NL4DV's response in JavaScript, \textsc{TouchPlot} is modified to support the required speech interactions (Listing~\ref{listing:mmplot-js-parsing}).
In particular, we parse the \variable{taskMap} to detect and apply any filters requested as part of the query (lines 6-12).
Next, we check if the input query specifies a new scatterplot that can be rendered by the system and adjust the view mappings accordingly (lines 13-16).
This sequential parsing of \variable{taskMap} and \variable{visList} allows using speech to apply filters, specify new scatterplots, or do both with a single query (Figure~\ref{fig:example-mmplot}).
Unlike previous examples, since this application uses D3 (as opposed to Vega-Lite) to create the visualization, when parsing NL4DV's output, we perform an added step of invoking the D3 code required to update the view (line 17).

\begin{listing}[t!]
\begin{minted}
[
baselinestretch=1,
fontsize=\scriptsize,
xleftmargin=15pt,
% xleftmargin=2.5pt,
linenos,
breaklines,
style=manni
]
{javascript}

$.post("/analyzeQuery", {"query": query})
  .done(function (responseString) {
    let nl4dvResponse = JSON.parse(responseString);
    let taskMap = nl4dvResponse['taskMap'],
        visList = nl4dvResponse['visList'];
    if("filter" in taskMap){ // query includes a filter
      for(let taskObj of taskMap['filter']){
        for(let attr of taskObj['attributes']){
          // use the attribute type, 'operator', and 'values' to apply requested filters
        }
      }
    }
    if(visList.length>0){ // query specifies a new chart
      let newVisSpec = visList[0]['vlSpec'];
      // check if newVisSpec is a scatterplot configuration supported by the system and modify the attribute-encoding mappings
    }
    // invoke the D3 code to update the view
  });
  
\end{minted}
\caption{\textcolor{black}{JavaScript code to parse NL4DV's output for supporting speech and multimodal interactions in \textsc{MMPlot} (Figure~\ref{fig:example-mmplot}-bottom).\vspace{-1.5em}}}
\label{listing:mmplot-js-parsing}
\vspace{-1.5em}
\end{listing}


\vspace{1em}
\noindent{}Implementing the aforementioned examples (NL-based Vega-Lite editor, DataTone's ambiguity widgets, \textsc{MMPlot}) would typically require developers to write hundreds of lines of code (in addition to the front-end code) requiring both NLP and visualization design knowledge (Figure~\ref{fig:overview}).
As illustrated above, with NL4DV, developers can accomplish the desired NLI capabilities with a single call to \function{analyze\_query()} and a few additional lines of code to parse NL4DV's response, enabling them to focus on the interface design and user experience.
\section{Discussion \add{and Future Work}}


\subsection{\add{Evaluation}}

\add{
In this paper, we illustrate NL4DV's query interpretation capabilities through sample queries executed on different datasets\footnote{Additional queries available at: \href{https://nl4dv.github.io/nl4dv/showcase.html}{\textcolor{urlblue}{https://nl4dv.github.io/nl4dv/showcase.html}}}.
As part of this initial validation, we used NL4DV to query tabular datasets containing 300-6000 rows and up to 27 attributes.
The toolkit's response time for these queries ranged between 1-18 sec.~(mean: 3 sec.)\footnote{Reported based on a MacBook Pro with a 6-core 2.9GHz processor and 16GB RAM running MacOS Catalina version 10.15.5}.
Besides sample queries, we also present applications employing NL4DV to highlight how the toolkit can reduce development viscosity and lower skill barriers (in terms of prior knowledge of NLP tools and techniques)~\cite{olsen2007evaluating}.
However, assessing the toolkit's usability and utility likely involves a more detailed evaluation in two ways.
First, we need to formally benchmark NL4DV's performance by executing it against a large corpus of NL queries.
To this end, an important area for future work is to collect labeled data on utterances people use to specify visualizations and use it to benchmark NL4DV and other visualization NLIs.
Second, we need to conduct a longitudinal study incorporating feedback from both visualization and NLP developers.
Going forward, we hope that the open-source nature of this research will help us conduct such a study in the wild, enabling us to assess NL4DV's practical usability, identify potential issues, and understand the breadth of possible applications.
}

\subsection{Supporting Follow-up Queries}

NL input presents the opportunity to support a richer visual analytic dialog through conversational interaction (as opposed to one-off utterances).
For example, instead of a single query including both visualization specification and filtering requests (e.g., ``\textit{Show a scatterplot of gross and budget highlighting only Action and Adventure movies}"), one can issue a shorter query to first gain an overview by specifying a visualization (e.g., ``\textit{Show a scatterplot of gross and budget}") and then issue a follow-up query to apply filters (e.g., ``\textit{Now just show Action and Adventure movies}").
However, supporting such a dialog through an interface-agnostic toolkit is challenging as it requires the toolkit to have context of system state in which the query was issued.
Furthermore, not all queries in a dialog may be follow-up queries.
While current systems like Evizeon~\cite{hoque2018applying} allow users to reset the canvas to reset the query context, explicitly specifying when context should be preserved/cleared while operating an interface-agnostic toolkit is impractical.

\begin{figure}[t!]
    \centering
    \includegraphics[width=\linewidth]{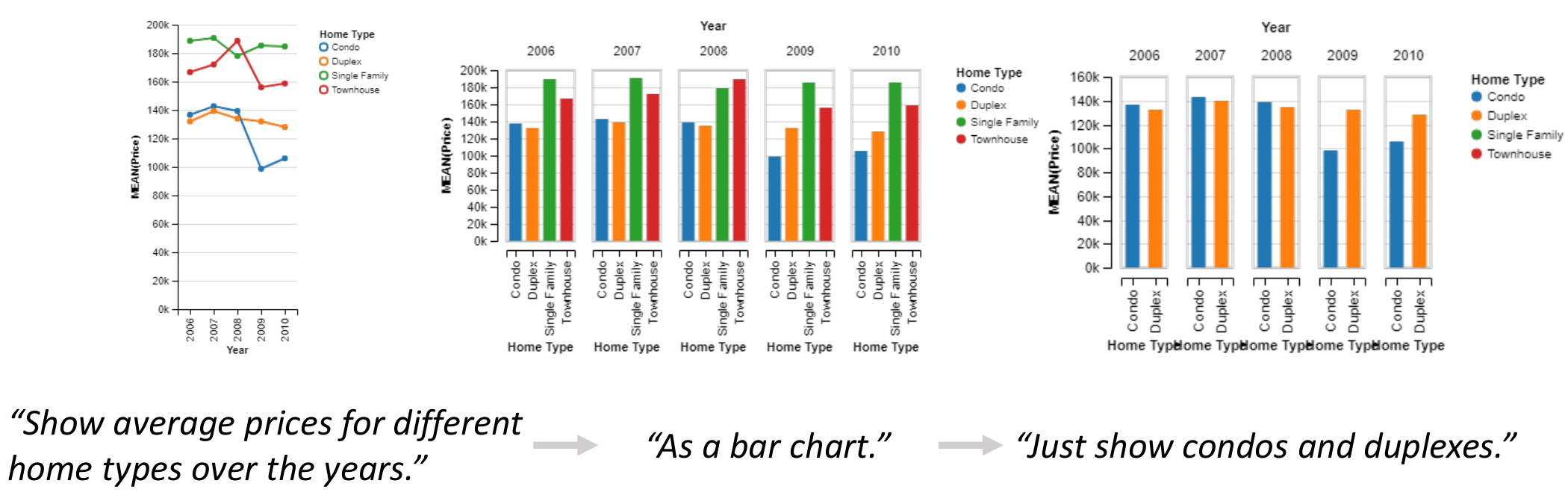}
    \vspace{-1.75em}
    \caption{Example of NL4DV supporting follow-up queries using the experimental \texttt{\textbf{dialog}} parameter. Here, the three visualizations are generated through consecutive calls to \function{render\_vis(\textit{query}, \textit{dialog}=true}).}
    \label{fig:follow-up}
    \vspace{-1.75em}
\end{figure}

NL4DV currently does not support follow-up queries.
However, as a first pass at addressing these challenges, we are experimenting with an additional \texttt{\textbf{dialog}} parameter to \function{analyze\_query()} and \function{render\_vis()} to support follow-up queries involving filtering and encoding changes.
Specifically, setting \texttt{\textbf{dialog}}$=$\texttt{true} notifies NL4DV to check for follow-up queries.
NL4DV uses conversational centering techniques~\cite{grosz1986attention,grosz1995centering} similar to prior visualization NLIs~\cite{hoque2018applying,srinivasan2018orko,srinivasan2020interweaving} to identify missing attributes, tasks, or visualization details in a query based on the toolkit's previous response.
Consider the example in Figure~\ref{fig:follow-up} showing queries issued in the context of a housing dataset.
In response to the first query ``\textit{Show average prices for different home types over the years,}" NL4DV generates a line chart by detecting the attributes \textit{Price}, \textit{House Type}, and \textit{Year}, and the task \textit{Derived Value} (with the operator \textit{AVG}).
Next, given the follow-up query ``\textit{As a bar chart,}" NL4DV infers the attributes and tasks from its previous response, updating the \texttt{mark} type and \texttt{encoding} channels in the Vega-Lite specification to create a grouped bar chart.
Lastly, with the third query, ``\textit{Just show condos and duplexes,}" detecting `condos' and `duplexes' as data values, NL4DV modifies the underlying \variable{taskMap} and applies a filter on the \textit{House Type} attribute.
Besides implementing additional parameters and functions to support conversational interaction in NL4DV, a general future research challenge is to investigate how interface context (e.g., active encodings, selections) can be modeled into a structured format that can be interpreted by interface-agnostic toolkits like NL4DV.

\subsection{\add{Improving Query Interpretation and Enabling Additional Query Types}}

\add{
Through our initial testing, we have already identified some areas for improvement in NL4DV's interpretation pipeline.
One of these is better inference of attribute types upon initialization.
To this end, we are looking into how we can augment NL4DV's data interpretation pipeline with recent semantic data type detection models that use both attribute names and values (e.g.,~\cite{zhang2019sato,hulsebos2019sherlock}).
Another area for improvement is task detection.
NL4DV currently leverages a combination of lexicon- and dependency-based approach to infer tasks.
Although this serves as a viable starting point, it is less reliable when there is uncertainty in the task keywords (e.g., the word ``\textit{relationship}" may not always map to a \textit{Correlation} task) or the keywords conflict with data attributes (e.g., the query ``\textit{Show the average cost of schools by region}" would currently apply a \textit{Derived Value} task on the \textit{Average Cost} attribute even though \textit{Average Cost} already represents a derived value).
As we collect more user queries, we are exploring ways to complement the current approach with semantic parsers (e.g.,~\cite{berant2013semantic,berant2014semantic}) and contemporary deep learning models (e.g.,~\cite{young2018recent,brown2020language}) that can infer tasks based on the query phrasing and structure.
Finally, a third area for improvement is to connect NL4DV to knowledge bases like WolframAlpha~\cite{wolframalpha} to semantically understand words in the input query.
Incorporating such connections will help resolve vague predicates for data values (e.g., `large', `expensive', `near')~\cite{setlur2016eviza,setlur2019inferencing} and may also reduce the need for developers to manually configure attribute aliases.}

\add{
Besides improving query interpretation, another theme for future work is to support additional query types.
As stated earlier, NL4DV is currently primarily geared to support visualization specification-oriented queries (e.g., ``\textit{What is the relationship between worldwide gross and content rating?}," ``\textit{Create a bar chart showing average profit by state}").
To aid development of full-fledged visualization systems, however, the toolkit needs to support a tighter coupling with an active visualization and enable other tasks such as question answering~\cite{kim2020answering} (e.g., ``\textit{How many movies grossed over 100M?}," ``\textit{When was the difference between the two stocks the highest?}") and formatting visualizations (e.g., ``\textit{Color SUVs green}," ``\textit{Highlight labels for countries with a population of more than 200M}").
Incorporating these new query types would entail making changes in terms of both the interpretation strategies (to identify new task categories and their parameters) and the output format (to include the computed ``answers" and changes to the view).
}

\subsection{Balancing Simplicity and Customization}

\add{NL4DV is currently targeted towards visualization developers who may not have a strong NLP background (\textbf{DG1}).
As such,}
NL4DV uses a number of default settings (e.g., preset rules for dependency parsing, empirically set thresholds for attribute detection) to minimize the learning curve and facilitate ease-of-use.
Developers can override some of these defaults (e.g., replace CoreNLP with spaCy~\cite{honnibal2017spacy}, adjust similarity matching thresholds) and also configure dataset-specific settings to improve query interpretation (e.g., attribute aliases, special words referring to data values, additional stopwords to ignore) (\textbf{DG4}).
Furthermore, by invoking \function{analyze\_query()} with the \param{debug} parameter set to \textit{true}, developers can also get additional details such as why an attribute was detected (e.g., semantic vs.~syntactic match along with the match score) or how a chart was implicitly inferred (e.g., using attributes vs.~attributes and tasks).

NLP or visualization experts, however, may prefer
\add{using custom modules for query processing or visualization recommendation.
To this end, visualization developers can override the toolkit's default recommendation engine by using the inferred attributes and tasks as input to their custom modules (i.e., ignoring the \variable{visList}).
However, NL4DV currently does not support using custom NLP models for attribute and task inference (e.g., using word embedding techniques to detect synonyms or classification models to identify tasks).
Going forward, as we gather feedback on the types of customizations developers prefer, we hope to provide the option for developers to replace NL4DV's heuristic modules with contemporary ML models/techniques.
Given the end goal of aiding prototyping of visualization NLIs, a challenge in supporting this customization, however, is to ensure that the output from the custom models can be compiled into NL4DV's output specification or to modify NL4DV's specification to accommodate additional information (e.g., classification accuracy) generated by the custom models.}

\section{Conclusion}

We present NL4DV, a toolkit that supports prototyping visualization NLIs.
Given a dataset and a NL query, NL4DV generates a JSON-based analytic specification composing of attributes, tasks, and visualizations inferred from the query.
Through example applications, we show how developers can use this JSON response to create visualizations in Jupyter notebooks through NL, develop web-based visualization NLIs, and augment existing visualization tools with NL interaction.
We provide NL4DV and the example applications as open-source software (\systemURL) and hope these will serve as valuable resources to advance research on NLIs for data visualization.

\acknowledgments{
This work was supported in part by a National Science Foundation Grant IIS-1717111.
}

\bibliographystyle{abbrv-doi}

\bibliography{template}
\end{document}